\documentclass[10pt]{article}
\usepackage{graphicx}
\usepackage{dcolumn}
\usepackage{bm}
\usepackage{fixmath}
\usepackage{amssymb,amsthm,amsmath}
\usepackage{xcolor,paralist,hyperref,titlesec,fancyhdr,etoolbox}
\usepackage[margin=2.54cm]{geometry}
\usepackage{authblk}
\usepackage{indentfirst,csquotes}

\DeclareMathOperator{\csch}{csch} 

\begin{document}

\title{\textbf{Applying the convolution theorem to the two-dimensional magnetic inverse problem}}
\author[]{M. T. M. Woodley\footnote{mw2970@bath.ac.uk}}
\affil[]{\small\textit{Department of Physics, University of Bath, Bath, BA2 7AY, UK.}}
\date{1$^{\mathrm{st}}$ September 2026}

\maketitle

\begin{abstract}
\noindent A collection and clarification is given of the key mathematical details of the method of applying the convolution theorem as part of the solution to the two-dimensional magnetic inverse problem: reconstructing current density from magnetic field data. Particular emphasis is given to obtaining a closed-form solution for the Fourier transform of the convolution kernel that features in the Biot-Savart law. Having done this, the Fourier-transformed current density is obtained, via the convolution theorem, as a function of Fourier-transformed magnetic field data and other variables.
\end{abstract}

\section{Introduction}
\noindent Inverse problems are very widespread throughout the mathematical sciences, with fields of application ranging from medical imaging \cite{Bertero-2006,Greensite-2004} to geophysics \cite{Blakely-1995,Richter-2020}. In contrast to a forward problem, in which data are calculated from model parameters, an inverse problem is characterised by inferring model parameters from data. Physically, this corresponds to reconstructing causes from observed effects. A key challenge to be confronted is the fact that inverse problems are often `ill-posed', rather than `well-posed'; these terms were introduced by J. Hadamard \cite{Hadamard-1902}. A problem may be said to be `well-posed' if it obeys the following three conditions:  

\begin{enumerate}
    \item The solution to the problem exists;
    \item The solution is unique;
    \item The solution is stable with respect to changes in other conditions upon which the problem may depend.
\end{enumerate}

\noindent The precise formulation of each of these conditions depends on mathematical context: for example, the space in which a solution is defined (uniqueness) or its associated topology (stability). If a solution fails to meet any of these criteria, then its associated problem is referred to as being ill-posed. 

The magnetic inverse problem is the reconstruction of current density from magnetic field data \cite{Roth-1989,Lima-2006,Grant-1994,Brandt-1995,Jooss-1998,Zuber-2018,Woodley-2025}. This has been of interest for several decades, particularly in conjunction with magnetometry investigations using superconducting quantum interference devices and, more recently, optically pumped magnetometers, due to the potential of both for extremely sensitive measurements (achieving noise densities of the order of $\mathrm{fT}/\sqrt{\mathrm{Hz}}$) \cite{Buchner-2018,Colombo-2016,Coussens-2024}. Such sensitive detection is of interest for a variety of applications, including the direct measurement of magnetic fields due to brain activity \cite{Coussens-2024,Brookes-2022}, and the non-invasive detection of defects in batteries, for example \cite{Bason-2022,Evans-2025,Brauchle-2023}.

Despite the variety of literature on the subject, presentations of the solution to the magnetic inverse problem can be rather compressed, and may omit certain steps, presumably because they are either deemed to be obvious or not sufficiently relevant to the focus of the article in question. Consequently, the key point of this article is to provide a reasonably complete overview of the method of solution to the magnetic inverse problem (when a unique solution is possible). In other words, its main function is to bring together several otherwise relatively disconnected results in a reasonably unified and pedagogical presentation, particularly for physicists. As such, these topics are presented for an audience interested in the magnetic inverse problem, but not necessarily familiar with the machinery of convolutions or Green functions more generally. The article is therefore written in a more expositional style than is typical, with a view to making the key logical steps explicit and approachable for a reasonably wide-ranging audience, in a presentation that is relatively self-contained. Consequently, no attempt has been made to compress the writing style. References to other literature for further reading on particular topics are provided throughout.

Of the above three criteria for a well-posed problem, the requirement for uniqueness is the one that will be addressed here in the context of the magnetic inverse problem. As a preliminary to solving the inverse problem, the associated magnetic forward problem is defined next.  

\subsection{The Biot-Savart law for volumes of current}
\noindent The Biot-Savart law describes the generation of a magnetic field by some bounded current, such that time-dependence does not need to be considered (i.e., a magnetostatic situation). In the case of current flowing through a wire, this law may be expressed in terms of an integral along some path that models the wire. When it is more appropriate to consider a volume of current, the Biot-Savart law instead takes the form of a triple integral \cite{Roth-1989,Woodley-2025}:

\begin{equation}\label{eq:B-S}
    \mathbold{H(r)} = \frac{1}{4\pi}\iiint\limits_{\Omega}\frac{\mathbold{J(r')}\times(\mathbold{r}-\mathbold{r'})}{|\mathbold{r}-\mathbold{r'}|^{3}}d^{3}\mathbold{r'},
\end{equation}

\noindent where $\mathbold{H(r)}$ is the magnetic field, $\mathbold{J(r')}$ is the current density, $\mathbold{r}$ is the position vector of the measurement of the magnetic field, and $\mathbold{r'}$ is the position vector associated with the current density. All these vectors are three-dimensional and have real components (i.e., they are in $\mathbb{R}^{3}$). $\Omega\subset\mathbb{R}^{3}$ denotes the bounded current-carrying region. The magnetic inverse problem amounts to solving \eqref{eq:B-S} for $\mathbold{J}$ in terms of $\mathbold{H}$.  

\section{Overview of the method of solution}\label{sec:Overview}
\noindent Some initial observations about \eqref{eq:B-S} may be made: 

\begin{enumerate}
    \item The cross product associates two components of $\mathbold{J}$ with each component of $\mathbold{H}$; calculating $\mathbold{J}$ from $\mathbold{H}$ therefore suggests a non-unique solution, making the problem ill-posed in general. This is apart from the fact that the integral also allows for different current densities to be integrated to yield the same magnetic field -- particularly so-called `silent' current sources \cite{Lima-2006}, which result in zero field; this is a second origin of non-uniqueness.
    \item The integral suggests that representing \eqref{eq:B-S} in terms of an integral transform may simplify its expression and facilitate solving for $\mathbold{J}$.
\end{enumerate}

\noindent Addressing these points, the two key steps in inverting the Biot-Savart law for the current density, $\mathbold{J}$, are:

\begin{enumerate}
    \item Reduce the number of dimensions of $\mathbold{J}$ to no more than two. Here, this means neglecting the $z$-component of $\mathbold{J}$, along with any $z$-dependence of the $x$- and $y$-components. This allows for mappings to be established between the $x$- and $y$- components of $\mathbold{H}$ and the $y$- and $x$- components of $\mathbold{J}$, respectively. The possibility of non-uniqueness due to `silent' sources is shown to be ruled out for this geometry in Section \ref{sect:Soln}. Note that, when combined, the $x$- and $y$-components of $\mathbold{J}$ also determine the $z$-component of $\mathbold{H}$ in the forward problem, but this will not be considered here.  
    \item Use Fourier transforms and the convolution theorem to transform the Biot-Savart integral \eqref{eq:B-S} into a product, in order to easily invert it for $\mathbold{J}$. The background and details of this transformation constitute the bulk of the remainder of this article.
\end{enumerate}

\noindent Prior to describing the convolution theorem, it is helpful to set it into a wider context, in order to introduce the mathematical objects that it involves.

\newpage

\section{Green functions and convolution kernels}
\noindent We start by introducing the concept of a Green function. Consider the following equation:

\begin{equation}\label{eq:L}
    T(x)f(x) = g(x),
\end{equation}

\noindent where $T$ is a linear operator (i.e., one that obeys $T(\alpha f_{1}(x_{1}) + \beta f_{2}(x_{2})) = \alpha T f_{1}(x_{1}) + \beta T f_{2}(x_{2})$, where $\alpha$ and $\beta$ are constants), $f$ and $g$ are functions, with $g$ denoted as a source, and $x\in\mathbb{R}^{n}$. Note that, by definition, $T(x)f(x) := (Tf)(x)$. A Green function is a mathematical object, $G = G(x,x')$, that solves the following equation \cite{Myint-U-2007}:

\begin{equation}\label{eq:Green}
    T(x)G(x,x') = \delta(x-x'),
\end{equation}

\noindent where $\delta(x-x')$ is the Dirac delta, or unit impulse (note that the word `function' is not being used here in connection with the Dirac delta, since its behaviour cannot be adequately described using functions, although distributions may be used instead; Green functions may also be treated rigorously in terms of distributions). A Green function is useful here because it is a means for calculating the solution $f(x)$ in \eqref{eq:L}. It is the translation (or sifting) property of the Dirac delta that is relevant to this:

\begin{equation}
    \int g(x')\delta(x-x')dx' = g(x),
\end{equation}

\noindent since, in combination with \eqref{eq:Green}, and the fact that $T$ is a linear function of $x$ only, we have

\begin{equation}
    \int g(x') T(x)G(x,x')dx' = T(x)\int G(x,x')g(x')dx' = g(x).
\end{equation}

\noindent Consequently, by comparing with \eqref{eq:Green}, we have an expression for the function $f$ in terms of the source, $g$, and the Green function, $G$ \cite{Myint-U-2007,Blackledge-2006}:

\begin{equation}\label{eq:f}
    f(x) = \int G(x,x')g(x')dx'.
\end{equation}

\noindent A comment next needs to be made about how $T$ may behave under coordinate translation.

\subsection{Translation equivariance and invariance}
\noindent Here, an operator is said to have translation equivariance if the combined result of applying the operator and a coordinate translation is indifferent to the order in which this is performed -- i.e., the operator in question, $T$, is generally affected by the shift operator, $S_{a}$, but also commutes with it, for all constant shifts of $a\in\mathbb{R}^{n}$:

\begin{equation}\label{eq:equiv}
    S_{a}T = TS_{a}.    
\end{equation}

\noindent A special case of this is translation invariance, whereby the operator $T$ is unaffected by coordinate translation -- i.e., it still commutes with the shift operator, but in a trivial sense: $S_{a}T = TS_{a} = T$. This nomenclature is consistent with the concept of an equivariant map in the context of group theory \cite{Pitts-2013}, and it allows for a precise distinction to be made between $T$ and its associated $G$.

\subsection{Translation equivariance and convolution kernels}
\noindent We are considering coordinate translations here because $\delta(x-x')$ is shifted by $x'$ and so, by \eqref{eq:Green}, so is $TG(x,x')$. Roughly speaking, if $T$ possesses translation equivariance (i.e., if it obeys \eqref{eq:equiv}), then it makes no difference whether this shift occurs before or after the application of $T$ to $G$. We may therefore apply the shift to $G$ before operating with $T$, which leads to $G(x,x') = K(x-x')$. More precisely, consider the unshifted cases: $G(x,0) := G_{0}(x)$ and $\delta(x-0) := \delta_{0}(x)$. In terms of the operators, from \eqref{eq:Green}, we have

\begin{equation}\label{eq:GreenZero}
    TG_{0} = \delta_{0}.
\end{equation}

\noindent Shifting by $x'$ gives

\begin{equation}\label{eq:shift}
    S_{x'}(TG_{0}) = S_{x'}\delta_{0}.
\end{equation}

\noindent By associativity of the operators and translation equivariance, \eqref{eq:equiv}, we have

\begin{equation}\label{eq:shift-2}
    S_{x'}(TG_{0}) = (S_{x'}T)G_{0} = (TS_{x'})G_{0} = T(S_{x'}G_{0}).
\end{equation}

\noindent Combining \eqref{eq:shift} and \eqref{eq:shift-2} and evaluating the operators at $x$ gives

\begin{equation}\label{eq:shift-x}
    (T(S_{x'}G_{0}))(x) = (S_{x'}\delta_{0})(x).
\end{equation}

\noindent We have 

\begin{equation}
    (T(S_{x'}G_{0}))(x) := T(x)(S_{x'}G_{0})(x) = T(x)G(x-x')
\end{equation}

\noindent and

\begin{equation}
    (S_{x'}\delta_{0})(x) = \delta_{0}(x-x') = \delta(x-x'),
\end{equation}

\noindent so \eqref{eq:shift-x} becomes

\begin{equation}
    T(x)G(x-x') = \delta(x-x').
\end{equation}

\noindent Comparing with \eqref{eq:Green} results in

\begin{equation}
    T(x)G(x,x') = T(x)G(x-x').
\end{equation}

\noindent Assuming that $G$ is unique, we obtain

\begin{equation}\label{eq:G-final}
    G(x,x') = G(x-x') := K(x-x').
\end{equation}

\noindent The use of translation equivariance in \eqref{eq:shift-2} is the key step in the above reasoning. A subtlety that is worth noting is that \eqref{eq:G-final} is not asserting that a function of two arguments is the same as a function of one argument, but rather that, for all values of $x$ and $x'$, the values of $G(x,x')$ and $G(x-x')$ are equal. The symbol $G$ is reused here, now a function of $x-x'$, as a shorthand for $(S_{x'}G_{0})(x)$. This mild ambiguity in notation is then resolved by the definition of $K$, as distinct from $G$, which makes it explicit that we are henceforth talking about a different mathematical object from $G$, with only one argument $(x-x')$. The form that \eqref{eq:f} now takes is a type of integral transform known as a convolution:

\begin{equation}\label{eq:convKernel}
    f(x) = \int\limits_{-\infty}^{\infty}K(x - x')g(x')dx',
\end{equation}

\noindent and $K(x-x')$ is referred to as a convolution kernel. As an illustration of the effects of coordinate shifts, make the replacements $x'\rightarrow x'-a$ and $x\rightarrow x-a$. This results in

\begin{equation}\label{eq:convKernelShift}
    f(x - a) = \int\limits_{-\infty}^{\infty}K(x - x')g(x'-a)dx'.
\end{equation}

\noindent Consequently, a shift in the input, $g(x'-a)$, is associated with the same shift in the output, $f(x-a)$, thereby showing translation equivariance. On the other hand, the convolution kernel, $K(x-x')$, is unaffected by this shift, and is therefore translation-invariant. In this way, translation equivariance of the operator $T$, given by \eqref{eq:equiv}, is connected with translation invariance of its Green function, $G$, associated via \eqref{eq:Green}. 

It is worth commenting that, in some fields, such as the study of linear time-invariant systems \cite{Oppenheim-1997}, the term `translation-invariant' is used in a more general way that may obscure the equivariance aspect. In this case, it is important to remember that it is the internal properties of the system in question that are translation-invariant; these are described by the impulse response of that system, which is equivalent to the convolution kernel shown here. By contrast, the relationship under coordinate shifting between a given input to the system and the output with which it is associated is translation-equivariant.  

Intuitively, a convolution is an operation that computes the total `overlap' between a pair of functions, aggregated as one is shifted across the other. In the case of \eqref{eq:convKernel}, this operation may be expressed compactly as $K*g$. The convolution has a particularly useful property -- the convolution theorem -- which will be addressed next, before then being applied to solve the magnetic inverse problem.

\section{The convolution theorem in two dimensions}
\noindent We consider two spatial dimensions here, as this is the upper limit for the number of dimensions that $\mathbold{J}$ may occupy such that a unique solution to the magnetic inverse problem generally exists. The convolution theorem states the following: for two functions (in two variables, $x$ and $y$), $f_{1} = f_{1}(x,y)$ and $f_{2} = f_{2}(x,y)$, 

\begin{equation}\label{eq:convTheorem}
    \mathcal{F}(f_{1}**f_{2}) = \mathcal{F}(f_{1})\mathcal{F}(f_{2}),
\end{equation}

\noindent where $\mathcal{F}$ denotes the Fourier transform \cite{Hecht-2017}. In other words, by transforming to Fourier space, a convolution may be expressed as a product. Note that ** is being used here, rather than *, to make it explicit that the convolution is performed in two dimensions. Consequently, if we have experimental data, $d$, that can be modelled as being given by a convolution,

\begin{equation}\label{eq:convData}
    d \sim f_{1}**f_{2},
\end{equation}

\noindent and we wish to reconstruct either $f_{1}$ or $f_{2}$, with the other function known, we can take the Fourier transform of \eqref{eq:convData}, apply \eqref{eq:convTheorem}, rearrange for the desired transformed function, and then take the inverse Fourier transform. This is the procedure that will be followed in order to solve the magnetic inverse problem.

\subsection{Definitions of the Fourier transform}
\noindent There are several slightly different ways to define the Fourier transform, depending on whether or not angular variables are used and, if so, on how angular prefactors (i.e., functions of $2\pi$) are allocated between the forward and inverse transforms. This can result in \eqref{eq:convTheorem} needing to be slightly modified. These definitions may be found in Ref.~\cite{Bracewell-1999}, for example.

\subsubsection{Definition 1: non-angular variables}
\noindent In this case, the transform pair (of forward and inverse transforms) is defined as

\begin{equation}
    \begin{split}
     \hat{f}(\xi_{x},\xi_{y}) = \mathcal{F}(f(x,y)) = \int\limits_{-\infty}^{\infty}\int\limits_{-\infty}^{\infty}f(x,y)\exp(-2\pi i(\xi_{x}x+\xi_{y}y))dxdy, \\
     f(x,y) = \mathcal{F}^{-1}(\mathcal{F}(f(x,y))) = \int\limits_{-\infty}^{\infty}\int\limits_{-\infty}^{\infty}\hat{f}(\xi_{x},\xi_{y})\exp(2\pi i(\xi_{x}x+\xi_{y}y))d\xi_{x}d\xi_{y}.
     \end{split}
\end{equation}

\noindent Here, $\xi_{x,y} = 1/\lambda_{x,y}$ are spatial frequencies in the $x$- and $y$-directions, associated with wavelengths $\lambda_{x,y}$. The lack of angular prefactors makes this definition convenient and symmetric between the forward and inverse transforms; it is often adopted in mathematics.

\newpage

\subsubsection{Definition 2: angular variables, symmetric prefactors}
\noindent Here, the transform pair is given by

\begin{equation}
    \begin{split}
     \hat{f}(k_{x},k_{y}) = \mathcal{F}(f(x,y)) = \frac{1}{2\pi}\int\limits_{-\infty}^{\infty}\int\limits_{-\infty}^{\infty}f(x,y)\exp(-i(k_{x}x+k_{y}y))dxdy, \\
     f(x,y) = \mathcal{F}^{-1}(\mathcal{F}(f(x,y))) = \frac{1}{2\pi}\int\limits_{-\infty}^{\infty}\int\limits_{-\infty}^{\infty}\hat{f}(k_{x},k_{y})\exp(i(k_{x}x+k_{y}y))dk_{x}dk_{y}.
     \end{split}
\end{equation}

\noindent Here, $k_{x,y} = 2\pi\xi_{x,y}$ are angular spatial frequencies. The prefactor in the forward transform of this definition means that the left-hand side of \eqref{eq:convTheorem} needs to be multiplied by $1/(2\pi)$ in order to preserve the convolution theorem.

\subsubsection{Definition 3: angular variables, non-symmetric prefactors}
\noindent Finally, we have the transform pair

\begin{equation}\label{eq:def3}
    \begin{split}
     \hat{f}(k_{x},k_{y}) = \mathcal{F}(f(x,y)) = \int\limits_{-\infty}^{\infty}\int\limits_{-\infty}^{\infty}f(x,y)\exp(-i(k_{x}x+k_{y}y))dxdy, \\
     f(x,y) = \mathcal{F}^{-1}(\mathcal{F}(f(x,y))) = \frac{1}{(2\pi)^{2}}\int\limits_{-\infty}^{\infty}\int\limits_{-\infty}^{\infty}\hat{f}(k_{x},k_{y})\exp(i(k_{x}x+k_{y}y))dk_{x}dk_{y}.
     \end{split}
\end{equation}

\noindent We will proceed using this definition, since angular variables are common in physics, and the lack of a prefactor in the forward transform preserves the convolution theorem as stated in \eqref{eq:convTheorem}.

\subsection{Proof of the convolution theorem in two dimensions}

\subsubsection{Theorem}
\noindent For two functions, $f_{1}(x,y)$ and $f_{2}(x,y)$, with respective Fourier transforms $\mathcal{F}(f_{1})(k_{x}, k_{y})$ and $\mathcal{F}(f_{2})(k_{x}, k_{y})$ (as defined by \eqref{eq:def3}), the following result holds:

\begin{equation}
    \mathcal{F}(f_{1}**f_{2}) = \mathcal{F}(f_{1})\mathcal{F}(f_{2}).
\end{equation}

\subsubsection{Proof}
\noindent By the definition of convolution in two dimensions,

\begin{equation}
    (f_{1}**f_{2})(x,y) = \int\limits_{-\infty}^{\infty}\int\limits_{-\infty}^{\infty}f_{1}(x-x',y-y')f_{2}(x',y')dx'dy'.
\end{equation}

\noindent By the definition of the Fourier transform in \eqref{eq:def3},

\begin{equation}
    \mathcal{F}(f_{1}**f_{2}) = \int\limits_{-\infty}^{\infty}\int\limits_{-\infty}^{\infty}\left(\int\limits_{-\infty}^{\infty}\int\limits_{-\infty}^{\infty}f_{1}(x-x',y-y')f_{2}(x',y')dx'dy'\right)\exp(-i(k_{x}x+k_{y}y))dxdy.
\end{equation}

\noindent Assuming that $f_{1}$ and $f_{2}$ obey Fubini's theorem, by changing the order of integration, we have

\begin{equation}
    \mathcal{F}(f_{1}**f_{2}) = \int\limits_{-\infty}^{\infty}\int\limits_{-\infty}^{\infty}f_{2}(x',y')\left(\int\limits_{-\infty}^{\infty}\int\limits_{-\infty}^{\infty}f_{1}(x-x',y-y')\exp(-i(k_{x}x+k_{y}y))dxdy\right)dx'dy'.
\end{equation}

\noindent Introducing the following change of variables: $u = x-x'$, $v = y-y'$, with $du = dx$, $dv = dy$ for the inner integral, results in

\begin{equation}
    \mathcal{F}(f_{1}**f_{2}) = \int\limits_{-\infty}^{\infty}\int\limits_{-\infty}^{\infty}f_{2}(x',y')\left(\int\limits_{-\infty}^{\infty}\int\limits_{-\infty}^{\infty}f_{1}(u,v)\exp(-i(k_{x}(u+x')+k_{y}(v+y')))dudv\right)dx'dy'.
\end{equation}

\noindent By splitting up the exponential terms and rearranging, we obtain

\begin{equation}
    \begin{split}
    \mathcal{F}(f_{1}**f_{2}) = \left(\int\limits_{-\infty}^{\infty}\int\limits_{-\infty}^{\infty}f_{1}(u,v)\exp(-i(k_{x}u+k_{y}v))dudv\right)\left(\int\limits_{-\infty}^{\infty}\int\limits_{-\infty}^{\infty}f_{2}(x',y')\exp(-i(k_{x}x'+k_{y}y'))dx'dy'\right) \\
    = \mathcal{F}(f_{1})\mathcal{F}(f_{2}),
    \end{split}
\end{equation}

\noindent again, using the definition of the Fourier transform. This completes the proof. Note that a similar proof may be demonstrated for the one-dimensional case \cite{Hecht-2017}.

\section{Applying the convolution theorem to the Biot-Savart law}
\noindent Having introduced the general concepts of Green functions, convolution kernels, and the convolution theorem, we now proceed to solving the magnetic inverse problem as a particular type of convolution. The key technical challenge in this case will be to obtain a closed-form expression for the Fourier transform of the associated Green function, as part of applying the convolution theorem. 

\subsection{Reducing the Biot-Savart law to a convolution in two dimensions}
\noindent We limit ourselves to the greatest dimensionality of $\mathbold{J}$ that allows for a general unique solution to the inverse problem: two. We do this by considering a sheet of current, extended in the $x$-$y$-plane, but sufficiently thin or otherwise structured in such a way that the $z$-component of $\mathbold{J}$ may be neglected, along with any $z$-dependence of the $x$- and $y$-components. Without loss of generality, we consider that this current sheet is centred at the origin, and that we are measuring its associated magnetic field at a measurement plane located at $z = z_{\mathrm{M}} > 0$. Under these assumptions, the $x$-component of the magnetic field, $H_{x}$ takes the following form \cite{Woodley-2025}:

\begin{equation}\label{eq:BS-reduction}
    H_{x}(x,y,z_{\mathrm{M}}) = \frac{1}{4\pi}\int\limits_{-\delta/2}^{\delta/2}\int\limits_{-\infty}^{\infty}\int\limits_{-\infty}^{\infty}\frac{J_{y}(x',y')\Delta z}{((x-x')^{2}+(y-y')^{2}+\Delta z^2)^{3/2}}dx'dy'dz'.
\end{equation}

\noindent Here, we have introduced $\Delta z = z_{\mathrm{M}} - z'$, where $z_{\mathrm{M}} > z'$, so that $\Delta z$ is the difference in $z$ between the measurement plane and the current element being integrated over. $H_{x}$ is considered here as an example, but an analogous relation to \eqref{eq:BS-reduction} associates $H_{y}$ with $-J_{x}$. Using Fubini's theorem for triple integrals, the order of integration may be changed in order to view this as a convolution in $x$ and $y$:

\begin{equation}
    H_{x}(x,y,z_{\mathrm{M}}) = (K ** J_{y})(x,y,z_{\mathrm{M}}) = \int\limits_{-\infty}^{\infty}\int\limits_{-\infty}^{\infty}K(x-x',y-y',z_{\mathrm{M}})J_{y}(x',y')dx'dy', 
\end{equation}

\noindent where the associated convolution kernel is

\begin{equation}\label{eq:K}
    K(x,y,z_{\mathrm{M}}) = \frac{1}{4\pi}\int\limits_{-\delta/2}^{\delta/2}\frac{\Delta z}{(x^{2}+y^{2}+\Delta z^{2})^{3/2}}dz'.
\end{equation}

\noindent Comparing with the general form of a convolution given in \eqref{eq:convKernel}, $H_{x}$ fulfils the role of the function $f$, and $J_{y}$ stands in for $g$ as the source.

\subsection{The Biot-Savart convolution kernel as a function of 1/\textit{r}}
\noindent Prior to taking the Fourier transform of \eqref{eq:K} in order to use the convolution theorem, it is useful to recognise that it may first be simplified: let $r = \sqrt{(x-x')^{2}+(y-y')^{2}+\Delta z^{2}}$. \eqref{eq:K} then becomes \cite{Jooss-1998,Zuber-2018,Woodley-2025}

\begin{equation}
    K(x,y,z_{\mathrm{M}}) = -\frac{1}{4\pi}\int\limits_{-\delta/2}^{\delta/2}\frac{\partial}{\partial\Delta z}\left(\frac{1}{r}\right)dz'.
\end{equation}

\noindent The Fourier transform (in $x$ and $y$) of this kernel then, assuming sufficient smoothness to permit Leibniz's integral rule, takes the form

\begin{equation}\label{eq:Fourier-kernel}
    \mathcal{F}(K)(k_{x}, k_{y}, z_{\mathrm{M}}) = -\frac{1}{4\pi}\int\limits_{-\delta/2}^{\delta/2}\frac{\partial}{\partial\Delta z}\mathcal{F}\left(\frac{1}{r}\right)dz'.
\end{equation}

\noindent Our task is therefore to calculate $\mathcal{F}(1/r)$.

\subsection{Comments on the validity of the Fourier transform in this case}
\noindent It should be noted here that a necessary condition for the validity of the Fourier transform of a function is that the integral of that function over its entire domain be finite, which does not hold for $\mathcal{F}(1/r)$ \cite{Bracewell-1999}. Two options for addressing this are:

\begin{enumerate}
    \item Using an alternative formulation not based on integrals (using parts of distribution theory).
    \item To recognise that the spatial derivative of this Fourier transform, which is what we are interested in calculating, is sufficiently well-defined for our purposes.
\end{enumerate}

\noindent At any rate, we will proceed assuming that the transform holds in a manner that is useful for physical modelling and we will not consider its validity any further here.

\section{The Hankel transform and its radial symmetry}
\noindent In considering $\mathcal{F}(1/r)$, we are talking about a two-dimensional Fourier transform of a radially symmetric function. This symmetry suggests transformation to polar coordinates -- see Ref.~\cite{Blakely-1995}, for example.

\subsection{Polar representation and the definition of the Hankel transform}
\noindent By \eqref{eq:def3}, we have

\begin{equation}\label{eq:F(1/r)}
    \mathcal{F}\left(\frac{1}{r}\right) = \int\limits_{-\infty}^{\infty}\int\limits_{-\infty}^{\infty}\frac{1}{\sqrt{(x-x')^{2}+(y-y')^{2}+\Delta z^{2}}}\exp(-i(k_{x}x+k_{y}y))dxdy.
\end{equation}

\noindent It is natural to express both the spatial coordinates and the spatial frequencies in a polar representation: let \\
\noindent $\rho = \sqrt{(x-x')^{2} + (y-y')^{2}}$, $x = \rho\cos(\phi)$, $y = \rho\sin(\phi)$, $dxdy = \rho d\rho d\phi$, $k = \sqrt{k_{x}^{2} + k_{y}^{2}}$, $k_{x} = k\cos(\varphi)$, and $k_{y} = k\sin(\varphi)$. \eqref{eq:F(1/r)} then becomes

\begin{equation}\label{eq:F(1/r)-rho}
    \begin{split}
    \mathcal{F}\left(\frac{1}{r}\right) = \int\limits_{0}^{2\pi}\int\limits_{0}^{\infty}\frac{1}{\sqrt{\rho^{2}+\Delta z^{2}}}\exp(-ik\rho\cos(\phi-\varphi))\rho d\rho d\phi \\
    = \int\limits_{0}^{\infty}\frac{1}{\sqrt{\rho^{2}+\Delta z^{2}}}\left(\int\limits_{0}^{2\pi}\exp(-ik\rho\cos(\phi-\varphi))d\phi\right)\rho d\rho.
    \end{split}
\end{equation}

\noindent The integrand of the inner integral is periodic over the limits of integration, so $\varphi$ may be set to zero, without loss of generality. To see this, let $\theta:=\phi-\varphi$, $d\theta = d\phi$, so that this inner integral becomes

\begin{equation}\label{eq:FPhi}
    \int\limits_{-\varphi}^{2\pi-\varphi}\exp(-ik\rho\cos(\theta))d\theta.
\end{equation}

\noindent Call \eqref{eq:FPhi} $F(\varphi)$. Using the fundamental theorem of calculus, we then have

\begin{equation}\label{eq:FConst}
    \begin{split}
    &\frac{dF}{d\varphi} = \frac{d}{d\varphi}\left(\int\limits_{-\varphi}^{2\pi-\varphi}\exp(-ik\rho\cos(\theta))d\theta\right) \\
    &= \exp(-i(k\rho\cos(2\pi-\varphi)))\frac{d}{d\varphi}(2\pi-\varphi) - \exp(-ik\rho\cos(-\varphi))\frac{d}{d\varphi}(-\varphi) \\
    &= \exp(-ik\rho\cos(-\varphi)) - \exp(-ik\rho\cos(2\pi-\varphi)) \\
    &= \exp(-ik\rho\cos(-\varphi)) - \exp(-ik\rho(\cos(2\pi)\cos(\phi)+\sin(2\pi)\sin(\phi)))\\
    &= \exp(-ik\rho\cos(\varphi))-\exp(-ik\rho\cos(\varphi))\\
    &= 0,
    \end{split}
\end{equation}

\noindent Consequently, $F(\varphi)$ is a constant, so $\varphi$ has no influence on this integral and may be neglected:

\begin{equation}
    \int\limits_{0}^{2\pi}\exp(-i(k\rho\cos(\phi-\varphi)))d\phi = \int\limits_{0}^{2\pi}\exp(-i(k\rho\cos(\phi)))d\phi.
\end{equation}

\noindent This result holds for the integral of any periodic function over its period \cite{Tolstov-1962}. The Fourier transform now takes the form

\begin{equation}\label{eq:Hankel-1}
    \mathcal{F}\left(\frac{1}{r}\right) = 2\pi\mathcal{F}_{0}\left(\frac{1}{r}\right),
\end{equation}

\noindent where 

\begin{equation}\label{eq:Hankel-2}
    \mathcal{F}_{0}\left(\frac{1}{r}\right) = \int\limits_{0}^{\infty}\frac{1}{\sqrt{\rho^{2}+\Delta z^{2}}}J_{0}(k\rho)\rho d\rho
\end{equation}

\noindent defines a zeroth-order Hankel transform. Here,

\begin{equation}\label{eq:J0}
    J_{0}(k\rho) = \frac{1}{2\pi}\int\limits_{0}^{2\pi}\exp(-ik\rho\cos(\phi))d\phi
\end{equation}

\noindent is a zeroth-order Bessel function of the first kind \cite{Blakely-1995}. The solution of \eqref{eq:Hankel-2} forms the final part of calculating the Fourier transform of the Biot-Savart kernel.

\newpage

\section{Solution to the magnetic inverse problem}\label{sect:Soln}

\subsection{Solution to \texorpdfstring{$\mathcal{F}_{0}(1/r)$}{}}
\noindent The solution to the Hankel transform in \eqref{eq:Hankel-2} is a known result, given by Gradshteyn and Ryzhik \cite{Gradshteyn-2014}, for example, as 

\begin{equation}\label{eq:exp}
    \mathcal{F}_{0}(1/r) = \frac{\exp(-k\Delta z)}{k},
\end{equation}

\noindent where $k\neq 0$ and $\Delta z > 0$. This a result that is frequently quoted without proof in literature on the magnetic inverse problem. However, a relatively direct way of proving that this result holds is outlined by Ref.~\cite{eyeballfrog-2022}. This involves using the fact that the Hankel transform is an involution, meaning that, for a function $f(r)$,

\begin{equation}\label{eq:invol-1}
    \mathcal{F}_{0}(\mathcal{F}_{0}(f(r))) = f(r).
\end{equation}

\noindent Consequently, in the case of $f(r) = 1/r$, where $r = \sqrt{\rho^{2}+\Delta z^{2}}$, it suffices to show that 

\begin{equation}\label{eq:invol-2}
    \mathcal{F}_{0}\left(\frac{\exp(-k\Delta z)}{k}\right) = \frac{1}{\sqrt{\rho^{2}+\Delta z^{2}}}.
\end{equation}

\noindent The left-hand side of \eqref{eq:invol-2} is

\begin{equation}\label{eq:Laplace-1}
     \int\limits_{0}^{\infty}\frac{\exp(-k\Delta z)}{k}J_{0}(\rho k)k dk = \frac{1}{\rho}\int\limits_{0}^{\infty}\exp(-\sigma q)J_{0}(q)dq,
\end{equation}

\noindent where we have introduced $q = \rho k$ and $\sigma = \Delta z/\rho$. Due to having previously re-expressed the two-dimensional Fourier transform in polar coordinates -- motivated by the radial symmetry of $1/r$ -- and the consequent integral over the non-negative half-line in \eqref{eq:Laplace-1} (rather than the entire real line), we have a Laplace transform, $\mathcal{L}$, of the Bessel function $J_{0}(q)$:

\begin{equation}\label{eq:laplace-2}
    \mathcal{L}(J_{0})(\sigma) = \int\limits_{0}^{\infty}J_{0}(q)\exp(-\sigma q)dq.
\end{equation}

\noindent Having noticed this, we can combine the properties of Laplace transforms and Bessel functions in order to extract an algebraic expression, from which we can obtain the right-hand side of \eqref{eq:invol-2}. A Bessel function of order $\nu$, $J_{\nu}(q)$, satisfies Bessel's equation:

\begin{equation}\label{eq:Bessel}
    q^{2}\frac{d^{2}J_{\nu}}{dq^{2}} + q\frac{dJ_{\nu}}{dq} + (q^{2} - \nu^{2})J_{\nu} = 0.
\end{equation}

\noindent In this particular case ($\nu = 0$), we have (for $q\neq 0$)

\begin{equation}\label{eq:Bessel-zero}
    q\frac{d^{2}J_{0}}{dq^{2}} + \frac{dJ_{0}}{dq} + qJ_{0} = 0.
\end{equation}

\noindent A key property of Laplace transforms is their ability to simplify equations by converting operations in calculus into algebraic operations; it is the algebraic structure of the Laplace-transformed version of \eqref{eq:Bessel-zero} that will demonstrate that \eqref{eq:invol-2} holds. We next consider the Laplace transform, $\mathcal{L}$, of each term in \eqref{eq:Bessel-zero}.  

\subsubsection{Transform 1}
\noindent For $\mathcal{L}(qJ_{0})$, we have 

\begin{equation}
    \mathcal{L}(qJ_{0}) = \int\limits_{0}^{\infty}qJ_{0}(q)\exp(-\sigma q)dq = -\int\limits_{0}^{\infty}J_{0}(q)\frac{\partial}{\partial \sigma}\exp(-\sigma q)dq.
\end{equation}

\noindent Via Leibniz's integral rule (for constant limits of integration), 

\begin{equation}
    \mathcal{L}(qJ_{0}) = -\frac{d}{d\sigma}\int\limits_{0}^{\infty}J_{0}(q)\exp(-\sigma q)dq = -\frac{d}{d\sigma}\mathcal{L}(J_{0}).
\end{equation}

\subsubsection{Transform 2}
\noindent Now consider $\mathcal{L}(\frac{dJ_{0}}{dq})$. We have

\begin{equation}\label{eq:Laplace-deriv}
    \mathcal{L}\left(\frac{dJ_{0}}{dq}\right) = \int\limits_{0}^{\infty}\frac{dJ_{0}}{dq}\exp(-\sigma q)dq.
\end{equation}

\noindent Via the product rule,

\begin{equation}
    \frac{\partial}{\partial q}(J_{0}\exp(-\sigma q)) = \frac{dJ_{0}}{dq}\exp(-\sigma q) - J_{0}\sigma\exp(-\sigma q),
\end{equation}

\noindent and \eqref{eq:Laplace-deriv} becomes

\begin{equation}
    \begin{split}
    &\mathcal{L}\left(\frac{dJ_{0}}{dq}\right) = \int\limits_{0}^{\infty}\frac{\partial}{\partial q}(J_{0}\exp(-\sigma q)) dq + \sigma\int\limits_{0}^{\infty}J_{0}\exp(-\sigma q)dq \\
    &= \lim_{t\rightarrow\infty}\left[J_{0}\exp(-\sigma q)\right]_{0}^{t} + \sigma \mathcal{L}(J_{0}) \\
    &= \sigma \mathcal{L}(J_{0}) - J_{0}(0) \\
    &= \sigma \mathcal{L}(J_{0}) - 1,
    \end{split}
\end{equation}

\noindent using $J_{0}(0) = \frac{1}{2\pi}\int\limits_{0}^{2\pi}\exp(0)d\phi = \frac{1}{2\pi}[\phi]_{0}^{2\pi} = 1$.

\subsubsection{Transform 3}
\noindent Finally, consider $\mathcal{L}(q\frac{d^{2}J_{0}}{dq^{2}})$:

\begin{equation}
    \mathcal{L}\left(q\frac{d^{2}J_{0}}{dq^{2}}\right) = \int\limits_{0}^{\infty}q\frac{d^{2}J_{0}}{dq^{2}}\exp(-\sigma q)dq.
\end{equation}

\noindent Similarly to Transform 1, via Leibniz's integral rule, we have

\begin{equation}
    \mathcal{L}\left(q\frac{d^{2}J_{0}}{dq^{2}}\right) = -\frac{d}{d\sigma}\int\limits_{0}^{\infty}\frac{d^{2}J_{0}}{dq^{2}}\exp(-\sigma q)dq = -\frac{d}{d\sigma}\mathcal{L}\left(\frac{d^{2}J_{0}}{dq^{2}}\right).
\end{equation}

\noindent Using $\frac{d^{2}}{dq^{2}} = \frac{d}{dq}(\frac{d}{dq})$, the reasoning for Transform 2, up to the penultimate line, may be reused for $\mathcal{L}\left(\frac{d^{2}J_{0}}{dq^{2}}\right)$ by making the substitution $J_{0}\rightarrow\frac{d J_{0}}{dq}$. This results in

\begin{equation}
    \begin{split}
    &\mathcal{L}\left(\frac{d^{2}J_{0}}{dq^{2}}\right) = \sigma\mathcal{L}\left(\frac{dJ_{0}}{dq}\right) - \frac{dJ_{0}}{dq}(0) \\
    &= \sigma (\sigma \mathcal{L}(J_{0}) - J_{0}(0)) - \frac{dJ_{0}}{dq}(0) \\
    &= \sigma^{2}\mathcal{L}(J_{0}) - \sigma.
    \end{split}
\end{equation}

\noindent This uses $\frac{dJ_{0}}{dq}(q) = \frac{1}{2\pi}\int\limits_{0}^{2\pi}\frac{\partial}{\partial q}\exp(-iq\cos(\phi))d\phi = \frac{1}{2\pi}\int\limits_{0}^{2\pi}(-i\cos(\phi))\exp(-iq\cos(\phi))d\phi$; consequently, we have  
$\frac{dJ_{0}}{dq}(0) = -\frac{i}{2\pi}\int\limits_{0}^{2\pi}\cos(\phi)\exp(0)d\phi = -\frac{i}{2\pi}[\sin(\phi)]_{0}^{2\pi} = 0$.

\subsubsection{Assembling the solution}
\noindent The Laplace transform of \eqref{eq:Bessel-zero} is

\begin{equation}\label{eq:Bessel-zero-Lap}
    \begin{split}
    &\mathcal{L}\left(q\frac{d^{2}J_{0}}{dq^{2}}\right) + \mathcal{L}\left(\frac{dJ_{0}}{dq}\right) + \mathcal{L}(qJ_{0}) = -\frac{d}{d\sigma}(\sigma^{2}\mathcal{L}(J_{0}) - \sigma) + \sigma\mathcal{L}(J_{0}) - 1 - \frac{d}{d\sigma}(\mathcal{L}(J_{0})) \\
    &= -\frac{d}{d\sigma}(\sigma^{2}\mathcal{L}(J_{0})) + \sigma\mathcal{L}(J_{0})- \frac{d}{d\sigma}(\mathcal{L}(J_{0})) = 0.  
    \end{split}
\end{equation}

\noindent Consequently, via the product rule, we obtain the following ordinary differential equation:

\begin{equation}\label{eq:Lap-ODE}
    \frac{d}{d\sigma}(\mathcal{L}(J_{0})) + \frac{\sigma}{1 +\sigma^{2}}\mathcal{L}(J_{0}) = 0,
\end{equation}

\noindent which is solved by

\begin{equation}\label{eq:Lap-sol}
    \mathcal{L}(J_{0}) = \frac{C}{\sqrt{1 + \sigma^{2}}},
\end{equation}

\noindent where $C$ is a constant. It follows that $C = 1$ by considering the asymptotic behaviour of \eqref{eq:Lap-sol} for large $\sigma$. In this case, the right-hand side of \eqref{eq:Lap-sol} takes the form $\sim C/\sigma$. Due to the $\exp(-\sigma q)$ term in $\mathcal{L}(J_{0})$ (see \eqref{eq:laplace-2}), for large $\sigma$, this exponential decay acts as a sharp cut-off filter for contributions to $\mathcal{L}(J_{0})$ away from the origin. Consequently, the left-hand side of \eqref{eq:Lap-sol} is principally controlled by the behaviour of $J_{0}$ near $q = 0$. By Taylor-expanding $J_{0}$ in this neighbourhood (near $q = 0$), the leading-order contribution to the behaviour of $\mathcal{L}(J_{0})$ is given by $\sim J_{0}(0)/\sigma$. Since $J_{0}(0) = 1$, as given above, \eqref{eq:Lap-sol} gives $1/\sigma \sim C/\sigma$, so that $C = 1$. This follows from a result known as Watson's lemma; further details may be found in Refs.~\cite{Watson-1918,Bender-1978}, for example. Recalling \eqref{eq:Laplace-1} and the substitution $\sigma = \Delta z / \rho$, we then have

\begin{equation}
    \mathcal{F}_{0}\left(\frac{\exp(-k\Delta z)}{k}\right) =\frac{1}{\rho}\mathcal{L}(J_{0}) = \frac{1}{\rho}\frac{1}{\sqrt{1 + (\frac{\Delta z}{\rho})^{2}}} = \frac{1}{\sqrt{\rho^{2}+\Delta z^{2}}},
\end{equation}

\noindent proving \eqref{eq:invol-2}, as required.

\subsection{Inverting for the current density}
\noindent Recalling \eqref{eq:Hankel-1}, we now have the solution to the two-dimensional Fourier transform of $1/r$:

\begin{equation}\label{eq:Fourier-sol}
    \mathcal{F}\left(\frac{1}{r}\right) = 2\pi\mathcal{F}_{0}\left(\frac{1}{r}\right) = 2\pi\frac{\exp(-k\Delta z)}{k},
\end{equation}

\noindent which was required in order to calculate the two-dimensional Fourier transform of the convolution kernel from the Biot-Savart law:

\begin{equation}
    \mathcal{F}(K) = -\frac{1}{4\pi}\int\limits_{-\delta/2}^{\delta/2}\frac{\partial}{\partial\Delta z}\mathcal{F}\left(\frac{1}{r}\right)dz'.
\end{equation}

\noindent Inserting \eqref{eq:Fourier-sol}, and remembering that $\Delta z := z_{\mathrm{M}} - z'$, we have

\begin{equation}\label{eq:BS-kernel-Fourier}
    \begin{split}
     &\mathcal{F}(K) = -\frac{1}{2}\int\limits_{-\delta/2}^{\delta/2}\frac{\partial}{\partial\Delta z}\left(\frac{\exp(-k\Delta z)}{k}\right)dz' \\
     &= \frac{1}{2}\int\limits_{-\delta/2}^{\delta/2}\exp(-k(z_{\mathrm{M}} - z'))dz' \\
     &= \frac{\exp(k\delta/2) - \exp(-k\delta/2)}{2k}\exp(-kz_{\mathrm{M}}) \\
     &= \frac{1}{k}\sinh\left(\frac{k\delta}{2}\right)\exp(-kz_{\mathrm{M}}),
    \end{split}
\end{equation}

\noindent where $k = \sqrt{k_{x}^{2} + k_{y}^{2}}$. Note that, for $k\delta/2 << 1$, \eqref{eq:BS-kernel-Fourier} reduces to

\begin{equation}
    \mathcal{F}(K) = \frac{\delta}{2}\exp(-kz_{\mathrm{M}}).
\end{equation}

\noindent Having obtained the Fourier transform of the kernel, $K$, We are now in a position to apply the convolution theorem, \eqref{eq:convTheorem}, to the Biot-Savart law, \eqref{eq:BS-reduction}, in order to re-express the latter to solve for current density, given magnetic field data. Restating the Biot-Savart law here for convenience, we have

\begin{equation}
    \begin{split}
    H_{x}(x,y,z_{\mathrm{M}}) = (K ** J_{y})(x,y,z_{\mathrm{M}}) = \int\limits_{-\infty}^{\infty}\int\limits_{-\infty}^{\infty}K(x-x',y-y',z_{\mathrm{M}})J_{y}(x',y')dx'dy', \\
    K(x,y,z_{\mathrm{M}}) = \frac{1}{4\pi}\int\limits_{-\delta/2}^{\delta/2}\frac{\Delta z}{(x^{2}+y^{2}+\Delta z^{2})^{3/2}}dz'.
    \end{split}
\end{equation}

\noindent Via the convolution theorem, \eqref{eq:convTheorem},

\begin{equation}\label{eq:convTheorem2}
    \mathcal{F}(H_{x}) = \mathcal{F}(K ** J_{y}) = \mathcal{F}(K)\mathcal{F}(J_{y}).    
\end{equation}

\noindent It is worth briefly raising the possibility of silent current sources, which result in no magnetic field, originally mentioned in Section \ref{sec:Overview}. By \eqref{eq:convTheorem}, for this two-dimensional, infinite-sheet geometry, a silent source, resulting here in $\mathcal{F}(H_{x}) = 0$, would require that $\mathcal{F}(K) = 0$ while $\mathcal{F}(J_{y}) \neq 0$. By \eqref{eq:BS-kernel-Fourier}, $\mathcal{F}(K) > 0$ for every finite $k > 0$, thereby ruling out silent sources, and their associated non-uniqueness, for this problem. Combining \eqref{eq:convTheorem2} and \eqref{eq:BS-kernel-Fourier}, we invert the magnetic forward problem, obtaining an expression for the current density, $J_{y}$, as required:

\begin{equation}\label{eq:final}
    J_{y} = \mathcal{F}^{-1}\left(\frac{\mathcal{F}(H_{x})}{\mathcal{F}(K)}\right) = \mathcal{F}^{-1}\left(k\csch\left(\frac{k\delta}{2}\right)\exp(kz_{\mathrm{M}})\mathcal{F}(H_{x})\right).
\end{equation}

\noindent A general characteristic of inverse problems may be appreciated here -- namely the amplification of noise. Here, in calculating $J_{y}$, any noise present in the magnetic field data, $H_{x}$, is amplified exponentially as a function of the wavenumber, $k$, and the distance of the measurement plane from the current-carrying volume, $z_{\mathrm{M}}$. Consequently, for practical current density reconstruction, the closer a magnetic field measurement may be performed to the source of that field, the better. This expression may be straightforwardly numerically implemented using a fast Fourier transform algorithm, such as in Ref.~\cite{Woodley-2025}, for example.

\newpage

\section{Conclusion}
\noindent The key mathematical details were described of using the convolution theorem to solve the two-dimensional magnetic inverse problem. This was intended as a means of `filling the gaps' for an audience interested in the magnetic inverse problem by bringing together the required derivations and results into a single, unified presentation. Significant parts of the derivations were devoted to calculating a closed-form solution to the two-dimensional Fourier transform of the convolution kernel that features in the Biot-Savart law, which defined the magnetic forward problem that was subsequently inverted. The requirement for numerical work, in addition to these derivations, was acknowledged in order to evaluate and practically implement \eqref{eq:final}.

\end{document}